\documentclass[english,12pt]{article}
\usepackage{amsmath}
\usepackage{amsfonts}
\usepackage{latexsym}
\usepackage{graphicx}
\usepackage{multicol}
\usepackage{color}
\usepackage[usenames,dvipsnames,svgnames,table]{xcolor}
\usepackage[utf8]{inputenc}
\usepackage{amsfonts}
\newcommand{\R}{\mbox{$ I \hspace{-1.2mm} R $}}
\begin{document}

\title{Multidimensional Scaling for Interval Data: INTERSCAL}

\maketitle

\markright{S. Winsberg, O. Rodr\'{\i}guez and E. Diday}

\author{S. Winsberg, O. Rodr\'{\i}guez and E. Diday \thanks{LISE--CEREMADE, Universit\'e de Paris IX Dauphine.  LISE--CEREMADE,
Universit\'e de Paris IX Dauphine.  IRCAM, 1 Place Igor Stravinsky, F--75004, Paris, France.}}
\date{\today}



\maketitle

\begin{abstract}
Standard multidimensional scaling takes as input a dissimilarity matrix of
general term $\delta _{ij}$ which is a numerical value. In this paper we
input $\delta _{ij}=[\underline{\delta _{ij}},\overline{\delta _{ij}}]$
where $\underline{\delta _{ij}}$ and $\overline{\delta _{ij}}$ are the lower
bound and the upper bound of the ``dissimilarity'' between the
stimulus/object $S_i$ and the stimulus/object $S_j$ respectively. As
output instead of representing each stimulus/object on a factorial plane by
a point, as in other multidimensional scaling methods, in the proposed
method each stimulus/object is visualized by a rectangle, in order to
represent dissimilarity variation. We generalize the classical scaling
method looking for a method that produces results similar to those obtained
by Tops Principal Components Analysis. Two examples are presented to
illustrate the effectiveness of the proposed method.
\end{abstract}

\section*{Keywords}
Symbolic object, multidimensional scaling, interval data, dissimilarity variation.

\pagebreak

\section{Introduction}

Let $S_1,S_2,\ldots ,S_m$ be $m$ stimuli/objects, we assume that the data
consist of a symmetric matrix $\Delta =([\delta _{ij}])=[\underline{\delta
_{ij}},\overline{\delta _{ij}}]$, $i,j=1,2,\ldots ,m$ where $[\underline{%
\delta _{ij}},\overline{\delta _{ij}}]$ represents the lower and upper
limits respectively of the set of the possible values for the
dissimilarity between the stimulus/object $S_i$ and the
stimulus/object $S_j$. The set of possible values for the dissimilarity
between the stimulus/object $S_i$ and the stimulus/object $S_j$ might
result from combining data from $N$ judges or sources, or alternatively it
might be a region of dissimilarity proposed by a single judge/source. Of
course, the interval of dissimilarities for each pair of stimuli/objects
could be trimmed on each end by five percent of the values, or the
inter--quartile range could be used instead of the entire interval. So
instead of a data table of dissimaliarities, that is a table of numerical
values for the dissimilarity of each stimulus/object pair, we have a data
table consisting of intervals representing the lower and upper limits of
dissimilarity for each stimulus/object pair.

Consider a set of stimuli consisting of rectangles of varying area, ie size,
and height--to--width ratio, ie shape, presented pairwise to subjects on a
computer \linebreak screeen. The subjects respond to each pair with a dissimilarity
judgment. The dissimilarity between stimulus $i$ and stimulus $j$ can be
represented as an interval, since each all the judges may not respond in the
same way. Therefore, each ``psychological", or ``perceptual", rectangle
is not precisely located, even though the corresponding ``physical"
rectangle occupies a point in two--dimensional space. Moreover, the
dimensions used to make the disssimilarity judgments might be size and
shape, or height and width, or something else. In fact an aim of
multidimensional scaling, MDS, is to locate the stimuli--objects in a
low--dimensional space, and interpret the dimensions giving rise to the
dissimilarity judgments. With our approach we will also determine how well
each stimulus/object is localized.

We have chosen to represent the objects $S_i$ and $S_j$ as hypercubes
in a low--dimensional Euclidean space, rather than say for example
hyperspheres because the hypercube representation is reflected
as a conjunction of $n$ properties, where $n$ is the dimensionality of the
space. This representation as a conjunction is appealing for two reasons.
The first reason is linguistic. In everyday language one refers to
rectangles having both an area lying between 10 and 20 square centimters,
and a height--to--width ratio lying between 0.5 and 0.8; one does not refer
to a rectangle with an area and shape centered at 15 square centimeters and
0.65 with a radius of 0.75, the radius to be expressed in just what units.
In fact when one queries a data base, the query is expressed as a
conjunction. The second reason is that this representation as a conjunction
fits within a data analytic framework concerning symbolic objects. This data
analytic framework has proved to be useful in dealing with large data bases.

A symbolic object is a model for an entity which can be an individual or a concept
of the real world equipped with a means of comparing the description of this
entity to the description of an individual observation. More precisely, it is
defined by: $i)$ a description $D$, say area $[10,20]$ and height--to--width
ratio $[0.5,0.8]$; $ii)$ a binary relation $R$, say $=$, $\leq $, or $\in$, permitting the comparison between
two descriptions to the entire set of descriptions $\cal D$; $iii)$ a function, or
mapping, $a$ providing a means of evaluating the result of the comparison
(using $R$) of the description of an individual in $\Omega$ the set of all
the individuals to the descripton
$D$. The extent of a symbolic object $a$ is the set of individuals
who fit the description. Thus,
a symbolic object is defined by the triple $s=(a,R,D)$ where $a$ depends on
the relation $R$ and the description $D$. Interval data are a type of
symbolic data, so we might have for example, $a(w)= [size(w) \in [10,20] ]
\wedge [ shape(w) \in [0.5,0.8]]$ defining a symbolic object $s=(a,R,D)$.

The representation of the
upper and lower values for the distance between symbolic object $S_i$ and
symbolic object $S_j$ is given in equations (1) and (2), below.

\textrm{Let be }$R_{S_i}$\textrm{\ the hypercube defined in }${\mbox{$
I \hspace{-1.2mm} R $}}^n$\textrm{\ by the symbolic object }$S_i$\textrm{, }
$R_{S_j}$\textrm{\ the hypercube in }${\mbox{$ I \hspace{-1.2mm} R $}}%
^n$\textrm{\ defined by the symbolic object }$S_j$\textrm{\ and let }$%
\underline{d_{ij}}$\textrm{\ end }$\overline{d_{ij}}$\textrm{\ the minimum
and the maximum Euclidean distances between }$R_{S_i}$\textrm{\ and }$R_{S_j}
$\textrm{. Then (where $\underline{x_{ik}}$ and $\overline{x_{jk}}$ are
defined in the equation (\ref{eq:eqcc41}) below) one could be represent
$\overline{d_{ij}}$ and $\underline{d_{ij}}$ as:}

\begin{equation}
\overline{d_{ij}}=\frac 12\sqrt{\sum\limits_{k=1}^n\left[ \left( \overline{%
x_{ik}}-\underline{x_{ik}}\right) +\left( \overline{x_{jk}}-\underline{x_{jk}%
}\right) +2\left| \frac{\overline{x_{ik}}-\underline{x_{ik}}}2-\frac{%
\overline{x_{jk}}-\underline{x_{jk}}}2\right| \right] ^2}  \label{eq:eq211}
\end{equation}
\begin{eqnarray}
\underline{d_{ij}}=\frac 14\sqrt{\left[ \sum\limits_{k=1}^n\left( \overline{%
x_{ik}}-\underline{x_{ik}}\right) +\left( \overline{x_{jk}}-\underline{x_{jk}%
}\right) -2\left| \frac{\overline{x_{ik}}-\underline{x_{ik}}}2-\frac{%
\overline{x_{jk}}-\underline{x_{jk}}}2\right| -\sim \right. }  \nonumber \\
\overline{\left. \sim \left| \left( \overline{x_{ik}}-\underline{x_{ik}}%
\right) +\left( \overline{x_{jk}}-\underline{x_{jk}}\right) -2\left| \frac{%
\overline{x_{ik}}-\underline{x_{ik}}}2-\frac{\overline{x_{jk}}-\underline{%
x_{jk}}}2\right| \right| \right] ^2}  \label{eq:eq212}
\end{eqnarray}

Den\oe ux and Masson (1999) have proposed a solution to this problem
minimizing the stress function, $\sigma(\mathcal{R})$, by gradient descent:

\[
\sigma (\mathcal{R})=\sum\limits_{i<j}(\underline{d_{ij}}-\underline{\delta
_{ij}})^2+\sum\limits_{i<j}(\overline{d_{ij}}-\overline{\delta _{ij}})^2,
\]

\noindent where $\underline{d_{ij}}$ and $\overline{d_{ij}}$ represent,
respectively, the minimum and the maximum Euclidean distances between of the
region $R_i$ and $R_j$ that they are looking for in ${\mbox{$ I
\hspace{-1.2mm} R $}}^p$ to represent the symbolic object $S_i$ and $S_j$
respectively. Their approach has two problems. First it could find a local
minimum. Second the minimization by gradient descent is not optimal and the
computation of the gradient of their stress function is difficult to
implement.

To avoid these problems we propose a method for multidimensional scaling of
interval data, INTERSCAL, with a different approach. We generalize the classical
scaling method of Torgenson (1958) and Gower (1966) by looking for a method
that produces results similar to the Tops Method in Principal Component
Analysis proposed by (Cazes, Chouakria, Diday and Schektman (1997)). The
Tops Method extends standard principal component analysis to interval data.
Standard principal component analysis, as a dimensional reduction method,
aims at reducing the number of descriptive feat\-ures while taking into
account the main structure of the data. In the Tops Method, similarly, the
aim is to reduce the number of interval features, called interval principal
components. The variability or the inaccuracy of the descriptive features
are expressed, after the reduction, by interval principal components and
visualized by rectangles in the factorial space.

\section{Multidimensional scaling for interval data}

In the Tops Method case the input are $m$ symbolic objects $S_1,S_2,\ldots ,S_m$
described by $n$ interval variables $X^1,X^2,\ldots ,X^n$ like we show in
equation (\ref{eq:eqcc41}).

\begin{equation}
\left(
\begin{array}{c}
S_1 \\
\vdots  \\
S_m
\end{array}
\right) =\left(
\begin{array}{ccc}
X_{S_11} & \cdots  & X_{S_1n} \\
\vdots  & \ddots  & \vdots  \\
X_{S_m1} & \cdots  & X_{S_mn}
\end{array}
\right) =\left(
\begin{array}{ccc}
\left[ \underline{x_{11}},\overline{x_{11}}\right]  & \cdots  & \left[
\underline{x_{1n}},\overline{x_{1n}}\right]  \\
\vdots  & \ddots  & \vdots  \\
\left[ \underline{x_{m1}},\overline{x_{m1}}\right]  & \cdots  & \left[
\underline{x_{mn}},\overline{x_{mn}}\right]
\end{array}
\right)   \label{eq:eqcc41}
\end{equation}

With this matrix we construct a new numerical matrix $M$ of $2^m\cdot n$
rows and $n$ columns as we show in the equation (\ref{eq:eqcc41}); then the
Tops Method applies the standard Principal Component Analysis method to the
matrix $M$.

\begin{equation}
M=\left[
\begin{array}{c}
\left[
\begin{array}{cccc}
\underline{x_{11}} & \underline{x_{12}} & \cdots & \underline{x_{1n}} \\
\underline{x_{11}} & \underline{x_{12}} & \cdots & \overline{x_{1n}} \\
\vdots & \vdots & \ddots & \vdots \\
\overline{x_{11}} & \overline{x_{12}} & \cdots & \overline{x_{1n}}
\end{array}
\right] \\
\left[
\begin{array}{cccc}
\underline{x_{21}} & \underline{x_{22}} & \cdots & \underline{x_{2n}} \\
\underline{x_{21}} & \underline{x_{22}} & \cdots & \overline{x_{2n}} \\
\vdots & \vdots & \ddots & \vdots \\
\overline{x_{21}} & \overline{x_{22}} & \cdots & \overline{x_{2n}}
\end{array}
\right] \\
\vdots \\
\left[
\begin{array}{cccc}
\underline{x_{m1}} & \underline{x_{m2}} & \cdots & \underline{x_{mn}} \\
\underline{x_{m1}} & \underline{x_{m2}} & \cdots & \overline{x_{mn}} \\
\vdots & \vdots & \ddots & \vdots \\
\overline{x_{m1}} & \overline{x_{m2}} & \cdots & \overline{x_{mn}}
\end{array}
\right]
\end{array}
\right] ,  \label{eq:eq27}
\end{equation}

Let $S_1,S_2,\ldots ,S_m$ be $m$ symbolic objects, we assume that the input
data consists of a symmetric matrix $\Delta $ defined by:

\begin{equation}
\Delta =\left[
\begin{array}{cccc}
\lbrack \underline{\delta _{11}},\overline{\delta _{11}}] & [\underline{%
\delta _{12}},\overline{\delta _{12}}] & \cdots & [\underline{\delta _{1m}},%
\overline{\delta _{1m}}] \\
\lbrack \underline{\delta _{21}},\overline{\delta _{21}}] & [\underline{%
\delta _{22}},\overline{\delta _{22}}] & \cdots & [\underline{\delta _{2m}},%
\overline{\delta _{2m}}] \\
\vdots & \vdots & \ddots & \vdots \\
\lbrack \underline{\delta _{m1}},\overline{\delta _{m1}}] & [\underline{%
\delta _{m2}},\overline{\delta _{m2}}] & \cdots & [\underline{\delta _{mm}},%
\overline{\delta _{mm}}]
\end{array}
\right] ,  \label{eq:eq26}
\end{equation}

\noindent where $\underline{\delta _{ij}}$ represents the lower bound of the
dissimilarity between the symbolic object $S_i$ and the symbolic object $%
S_j$ and $\overline{\delta _{ij}}$ represents the upper bound of the
dissimilarity between the symbolic object $S_i$ and the symbolic object $%
S_j$.

It is well known that there is a {\em duality property} between principal
components analysis and classical multidimensional scaling where the
dissimilarities are given by Euclidean distances. Formally, if $\mu _i$
and $\xi _i$ are the eigenvalues and eigenvectors of the principal
components analysis respectively for $i=1,2,\ldots ,n$, and we denote
by $\lambda _i$  and $v_i$ the eigenvalues and eigenvectors of
the multidimensional scaling respectively for $i=1,2,\ldots ,n$ then
$\mu _i=\lambda _i$ and $\xi _i=X^tv_i$ for $i=1,2,\ldots ,n$.

If we want to get a Symbolic Multidimensional Scaling method that has the
duality property with the Tops Principal Components Analysis method, when
dissimilarity is modeled by an Euclidean distance, we need as input the
dissimilarities between all the rows of the matrix $M$ defined in (\ref
{eq:eq27}), because the Tops Principal Component Analysis method starts by
doing a classical principal component analysis of the matrix $M$ defined in (%
\ref{eq:eq27}).

Since the size of $M$ has $m\cdot 2^n$ rows and $n$ columns, we should have
as input a matrix $\Delta $ of size $m\cdot 2^n\times m\cdot 2^n$ but
this is clearly impossible, because we only have two dissimilarities, that
is the maximun and the minimum, for each pair of symbolic objects.

So it is impossible to find a Symbolic Multidimensional Method that has a
duality property with the Tops Principal Component Analysis. Therefore we
will find an approximate solution.

Let:

\begin{equation}
\begin{array}{c}
\underline{\delta _{ij}}={\min_{x\in R_{S_i},\; y\in R_{S_j}} }d(x,y) \\
\overline{\delta _{ij}}={\max_{x\in R_{S_i},\; y\in R_{S_j}} }d(x,y)
\end{array}
,  \label{eq:eq28}
\end{equation}

\begin{figure}[h]
\begin{center}
\begin{picture}(61.73,32.80)
\put(4.59,11.23){\framebox(10.93,11.46)[ct]{$R_{S_i}$}}
\put(27.51,0.12){\framebox(16.75,17.46)[cc]{$R_{S_j}$}}
\put(15.52,15.05){\line(1,0){11.99}}
\multiput(4.59,22.69)(0.23,0.12){68}{\line(1,0){0.23}}
\multiput(44.27,0.12)(0.12,0.12){146}{\line(0,1){0.12}}
\multiput(61.73,17.58)(-0.38,0.12){109}{\line(-1,0){0.38}}
\put(11.76,15.05){\makebox(0,0)[cc]{$\alpha_i$}}
\put(30.40,15.05){\makebox(0,0)[cc]{$\alpha_j$}}
\put(0.82,23.69){\makebox(0,0)[cc]{$\beta_i$}}
\put(48.38,0.12){\makebox(0,0)[cc]{$\beta_j$}}
\put(36.51,30.40){\makebox(0,0)[cc]{$\overline{\delta _{ij}}$}}
\put(20.99,18.75){\makebox(0,0)[cc]{$\underline{\delta _{ij}}$}}
\end{picture}
\end{center}
\caption{Minimum and maximum distances between 2 hypercubes.}
\label{fig21}
\end{figure}
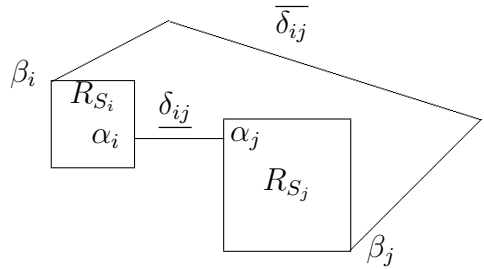

If we fix the hypercube $R_{S_i}$, it is clear that there are points $\alpha
_{ij}=(\alpha _1^{ij},\alpha _2^{ij},\ldots ,\alpha _n^{ij})$ $\in $ $R_{S_i}
$ and $\alpha _{ji}=(\alpha _1^{ji},\alpha _2^{ji},\ldots ,\alpha _n^{ji})$ $%
\in $ $R_{S_j}$, for $j=1,2,\ldots ,m$ such that $\underline{\delta _{ij}}%
=d(\alpha _{ij},\alpha _{ji})$. In a similar way there are points $%
\beta _{ij}=(\beta _1^{ij},\beta _2^{ij},\ldots ,\beta _n^{ij})\in R_{S_i}$
et $\beta _{ji}=\linebreak(\beta _1^{ji},\beta _2^{ji},\ldots ,\beta
_n^{ji})\in R_{S_j}$ such that $\overline{\delta _{ij}}=d(\beta
_{ij}^{},\beta _{ji})$ for $j=1,2,\ldots ,m,$ as we show in Figure
 \ref{fig21} for $n=2$. Since $j$ varies from $1$ to $m$ values, for each hypercube
 $R_{S_i}$ we have $m$ points $\alpha _{ij}$ and $m$ points $\beta _{ij}
$ and so we have $2mm$ dissimilarities (we take into account the maximum
and minimum dissimilarity between a hypercube and itself).
In the case where the MDS is used for data reduction of a matrix $X$ as
defined in equation (\ref{eq:eqcc41}), the mimimum dissimilarity between an object and
itself is zero, as derived from equation (\ref{eq:eq212}), whereas the maximum dissimilarity
may be derived from equation (\ref{eq:eqcc41}). In the case where the data consist of the
minimum and maximum dissimilarity for each object pair, both the minimum
and maximum dissimilarity between each object and itself are zero. In this
latter case most probably there are no data values for the dissimilarity
between an object and itself.
But, since $\underline{\delta _{ij}}=d(\alpha _{ij},\alpha _{ji})=%
\underline{\delta _{ji}}=d(\alpha _{ji},\alpha _{ij})$ et $\overline{\delta
_{ij}}=d(\beta _{ij},\beta _{ji})=\overline{\delta _{ji}}=d(\beta
_{ji},\beta _{ij})$ we have $2m+2(m-1)+\cdots +2=2\sum_{i=1}^mi=m(m+1)$
dissimilarities. These $m(m+1)$ dissimilarities include the 2m dissimilarities (maximum and
minimum) for each object and itself. When dealing with pairwise
dissimilarity data, since these 2m dissimilarities are zero, there are in
reality $m(m-1)$ dissimilarities, that is, $\frac{m(m-1)}{2}$ minimum and $\frac{m(m-1)}{2}$
maximum values, corresponding to the $\frac{m(m-1)}{2}$ pairs.

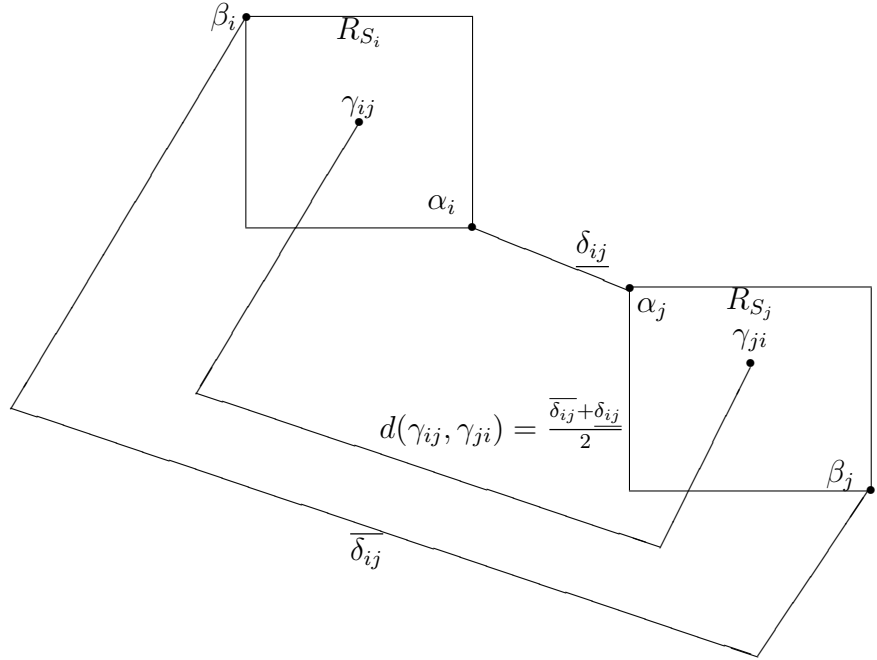
\begin{figure}[h]
\begin{center}
\unitlength=1.00mm \special{em:linewidth 0.4pt} \linethickness{0.4pt}
\begin{picture}(118.00,80.00)
\put(34.00,50.00){\framebox(30.00,28.00)[ct]{$R_{S_i}$}}
\put(85.00,15.00){\framebox(32.00,27.00)[ct]{$R_{S_j}$}}
\put(31.00,78.00){\makebox(0,0)[cc]{$\beta_i$}}
\put(60.00,53.00){\makebox(0,0)[cc]{$\alpha_i$}}
\put(64.00,50.00){\line(5,-2){21.00}}
\put(80.00,47.00){\makebox(0,0)[cc]{$\underline{\delta_{ij}}$}}
\put(34.00,78.00){\line(-3,-5){31.33}}
\put(2.67,26.00){\line(3,-1){99.33}}
\put(102.00,-7.00){\line(2,3){14.67}}
\put(50.00,7.00){\makebox(0,0)[cc]{$\overline{\delta_{ij}}$}}
\put(88.00,39.00){\makebox(0,0)[cc]{$\alpha_j$}}
\put(113.00,17.00){\makebox(0,0)[cc]{$\beta_j$}}
\put(49.00,64.00){\line(-3,-5){21.67}}
\put(27.33,28.00){\line(3,-1){61.67}}
\put(89.00,7.33){\line(1,2){12.00}}
\put(49.00,66.00){\makebox(0,0)[cc]{$\gamma _{ij}$}}
\put(101.00,35.00){\makebox(0,0)[cc]{$\gamma _{ji}$}}
\put(68.00,24.00){\makebox(0,0)[cc]{$d(\gamma_{ij},\gamma_{ji})= \frac{\overline{\delta_{ij}}+\underline{\delta _{ij}}}2$}}
\put(49.00,64.00){\circle*{1.00}}
\put(101.00,32.00){\circle*{1.00}}
\put(85.00,42.00){\circle*{1.00}}
\put(64.00,50.00){\circle*{1.00}}
\put(34.00,78.00){\circle*{1.00}}
\put(117.00,15.00){\circle*{1.00}}
\end{picture}
\end{center}
\caption{Mean distances between 2 hypercubes.}
\label{fig22}
\end{figure}

If we fix the hypercube $R_{S_i}$, there are also points $\gamma
_{ij}=(\gamma _1^{ij},\gamma _2^{ij},\ldots ,\gamma _n^{ij})$ $\in $ $%
R_{S_i} $ and $\gamma _{ji}=(\gamma _1^{ji},\gamma _2^{ji},\ldots ,\gamma
_n^{ji})$ $\in $ $R_{S_j}$, for $j=1,2,\ldots ,m$ such that $d(\gamma
_{ij},\gamma _{ji})=\frac{\overline{\delta _{ij}}+\underline{\delta _{ij}}}2$%
, as we show in Figure \ref{fig22}. This produces $m$ dissimilarities.

The idea, then, is to do a Multidimensional Scaling of the distance
matrix $\widetilde{\Delta }$ defined for the equation (%
\ref{eq:eq210}). For each hypercube $R_{S_i}$ the matrix $\widetilde{%
\Delta }$ has two rows, in the first row we use the minimum dissimilarity
and the maximum
dissimilarity among a hypercube and itself, whereas we use
the dissimilarity minimum and the average dissimilarity among each
different couple of hypercubes, that is to say, we use $2m$
dissimilarities. In the second row of the matrix
$\widetilde{\Delta }$ we use the maximum dissimilarity and the minimum dissimilarity
among a hypercube and itself, and we use the average dissimilarity and the
maximum dissimilarity among each different couple of hypecubes, in
this row we also use $2m$ dissimilarities, but as the average dissimilarities
were already employed we really use $m$
dissimilarities, therefore for each hypercube we use $3m$
dissimilarities. Then, since $d(x,y)=d(y,x)$, in total we use $%
3m+3(m-1)+\cdots +3=3\sum_{i=1}^mi=\frac 32m(m+1)>m(m+1)$
dissimilarities. Note that $\widetilde{\Delta }$ is a symetric matrix and
its size is $2m\times 2m$. Since for each hypercube $R_{S_i}$ we have
two rows, we can compute a principal coordinate minimum and
maximum, i.e. principal coordinates of interval type.

\begin{equation}
\widetilde{\Delta }=\left[
\begin{array}{ccccccccc}
0 & \overline{\delta _{11}} & \underline{\delta _{12}} & \frac{\overline{%
\delta _{12}}+\underline{\delta _{12}}}2 & \underline{\delta _{13}} & \frac{%
\overline{\delta _{13}}+\underline{\delta _{13}}}2 & \cdots & \underline{%
\delta _{1m}} & \frac{\overline{\delta _{1m}}+\underline{\delta _{1m}}}2 \\
\overline{\delta _{11}} & 0 & \frac{\overline{\delta _{12}}+\underline{%
\delta _{12}}}2 & \overline{\delta _{12}} & \frac{\overline{\delta _{13}}+%
\underline{\delta _{13}}}2 & \overline{\delta _{13}} & \cdots & \frac{%
\overline{\delta _{1m}}+\underline{\delta _{1m}}}2 & \overline{\delta _{1m}}
\\
\underline{\delta _{21}} & \frac{\overline{\delta _{21}}+\underline{\delta
_{21}}}2 & 0 & \overline{\delta _{22}} & \underline{\delta _{23}} & \frac{%
\overline{\delta _{23}}+\underline{\delta _{23}}}2 & \cdots & \underline{%
\delta _{2m}} & \frac{\overline{\delta _{2m}}+\underline{\delta _{2m}}}2 \\
\frac{\overline{\delta _{21}}+\underline{\delta _{21}}}2 & \overline{\delta
_{21}} & \overline{\delta _{22}} & 0 & \frac{\overline{\delta _{23}}+%
\underline{\delta _{23}}}2 & \overline{\delta _{23}} & \cdots & \frac{%
\overline{\delta _{2m}}+\underline{\delta _{2m}}}2 & \overline{\delta _{2m}}
\\
\underline{\delta _{31}} & \frac{\overline{\delta _{31}}+\underline{\delta
_{31}}}2 & \underline{\delta _{32}} & \frac{\overline{\delta _{32}}+%
\underline{\delta _{32}}}2 & 0 & \overline{\delta _{33}} & \cdots &
\underline{\delta _{3m}} & \frac{\overline{\delta _{3m}}+\underline{\delta
_{3m}}}2 \\
\frac{\overline{\delta _{31}}+\underline{\delta _{31}}}2 & \overline{\delta
_{31}} & \frac{\overline{\delta _{32}}+\underline{\delta _{32}}}2 &
\overline{\delta _{32}} & \overline{\delta _{33}} & 0 & \cdots & \frac{%
\overline{\delta _{3m}}+\underline{\delta _{3m}}}2 & \overline{\delta _{3m}}
\\
\vdots & \vdots & \vdots & \vdots & \vdots & \vdots & \ddots & \vdots &
\vdots \\
\underline{\delta _{m1}} & \frac{\overline{\delta _{m1}}+\underline{\delta
_{m1}}}2 & \underline{\delta _{m2}} & \frac{\overline{\delta _{m2}}+%
\underline{\delta _{m2}}}2 & \underline{\delta _{m3}} & \frac{\overline{%
\delta _{m3}}+\underline{\delta _{m3}}}2 & \cdots & 0 & \overline{\delta
_{mm}} \\
\frac{\overline{\delta _{m1}}+\underline{\delta _{m1}}}2 & \overline{\delta
_{m1}} & \frac{\overline{\delta _{m2}}+\underline{\delta _{m2}}}2 &
\overline{\delta _{m2}} & \frac{\overline{\delta _{m3}}+\underline{\delta
_{m3}}}2 & \overline{\delta _{m3}} & \cdots & \overline{\delta _{mm}} & 0
\end{array}
\right] .  \label{eq:eq210}
\end{equation}

\noindent \textsc{Algorithm for Multidimensional Scaling of interval--value
dissimilarity data}

\begin{description}
\item[Step 1:]  Obtain the dissimilarities $\left\{\left[ \underline{\delta
_{ij}},\overline{\delta _{ij}}\right] \right\}_{i,j=1,2,\ldots ,m}$.

\item[Step 2:]  Compute the matrix $\widetilde{\Delta }=(\widetilde{\delta }_{ij})_{i,j=1,2,\ldots ,2m}$
defined in the equation (\ref{eq:eq210}).

\item[Step 3:]  Find the matrix $B=\{[b_{ij}]\}_{i,j=1,2,\ldots ,2m}$:
\[
b_{ij}=-\frac 12\left( \widetilde{\delta }_{ij}^2-\frac
1{2m}\sum\limits_{r=1}^{2m}\widetilde{\delta }_{rj}^2-\frac
1{2m}\sum\limits_{s=1}^{2m}\widetilde{\delta }_{is}^2+\frac
1{(2m)^2}\sum\limits_{r=1}^{2m}\sum\limits_{s=1}^{2m}\widetilde{\delta }%
_{rs}^2\right)
\]

\item[Step 4:]  Find the eigenvalues $\lambda _1,\lambda _2,\ldots ,\lambda
_{2m}$ and the associated eigenvectors \linebreak $v_1,v_2,\ldots ,v_{2m}$ of $B$.

\item[Step 5:]  Compute the coordinates of the $2m$ points in ${\mbox{$ I
\hspace{-1.2mm} R $}}^n$ using the formula:
\[
x_{ri}=\sqrt{\lambda _r}v_{ir}\mbox{ for }r=1,2,\ldots ,2m\mbox{ and }%
i=1,2,\ldots ,n\mbox{.}
\]

\item[Step 6:]  Construct the principal coordinates of interval type $%
X_1^I,X_2^I,\ldots ,X_m^I$ from the numerical coordinates $X_1,X_2,\ldots
,X_{2m}$ ($X_i=(x_{i1},x_{i2},\ldots ,x_{in})$). Let $L_{S_i}$ be the set of
row numbers in the matrix $\widetilde{M}$ referring to the symbolic object $%
S_i$. It is clear that $L_{S_i}=\{2i-1,2i\}$. If $X_{S_ij}=[\underline{x_{ij}},%
\overline{x_{ij}}]$ is the value of the principal component of interval type
$X_j^I$ for the symbolic object $S_i$ then:

\[
\underline{x_{ij}}={\min_{k\in L_{S_i}}}(x_{kj})={\min_{k\in\{2i-1,2i\}}}%
(x_{kj}),
\]

\[
\overline{x_{ij}}={\max_{k\in L_{S_i}}}(x_{kj})={\max_{k\in \{2i-1,2i\}}}%
(x_{kj}).
\]
\end{description}

The solution for $X$ is not unique for $B=V \Lambda V^t = XTT^tX^t$ for any
$TT^t=I$. Any rigid rotation is an example of matrix of type $T$. We choose the
solution corresponding to principal axes. The first axis maximizes the variance
of the $\alpha _i,\beta _i$ $i=1,2,\ldots ,m$. However, since any rotation is
also a solution, one may wish to rotate the principal axes solution in order to
obtaing axes which are more interpretable.

Let us consider the special case when all the interval $\left[ \underline{%
\delta _{ij}},\overline{\delta _{ij}}\right] $ are trivial that is they are
points, i.e. $\underline{\delta _{ij}}=\overline{\delta _{ij}}=\delta _{ij}$%
, for $i=1,2,\ldots ,m$; $j=1,2,\ldots ,m$. Then $\frac{\underline{\delta
_{ij}}-\overline{\delta _{ij}}}2=0$ and $\frac{\underline{\delta _{ij}}+%
\overline{\delta _{ij}}}2=\delta _{ij}$. Define $\widetilde{B}$ to be the
scalar product matrix for the classical Torgerson--Gower scaling. Let be $%
\widetilde{B}=\widetilde{V}\widetilde{\lambda }\widetilde{V}$. Then, $%
\lambda _s=2\widetilde{\lambda }_s$ for $s=1,2,\ldots ,q$ where $q$ is the
number of strictly positive eigenvalues. Moreover $\widetilde{v}%
_{ir}=2v_{2i-1,r}=x_{2i-1,r}=\sqrt{2}v_{2i,r}=x_{2i,r}$ for $r=1,2,\ldots ,q$%
; $i=1,2,\ldots ,m$.

In this case then $\underline{x_{ir}}=\overline{x_{ir}}=x_{ir}$ for $%
i=1,2,\ldots ,m$ and $r=1,2,\ldots ,q$. And then in this special case
\texttt{INTERSCAL} and Torgerson--Gower scaling are equivalent. Therefore
Torgerson--Gower scaling is a spacial case of \texttt{INTERSCAL} occurring
when all of the observed dissimilarity intervals collapse to points.

INTERSCAL Multidimensional Scaling method has an advantage with respect to
the Tops Principal Component method. The size of the matrix where the
algorithm computes the eigenvalues and eigenvectors for the INTERSCAL
Multidimensional Scaling method is $2m\times 2m$, while in Tops Principal
Components Analysis method, it may be $m\cdot 2^n\times m\cdot 2^n$.

\section{Examples}

We have analyzed two data sets. First, a data set already explored in the
principal components context, and second a more traditional multidimensional
scaling data set involving judged dissimilarities. MDS may be used to analyze
proximity data or alternatively it may be used
for dimension reduction. In the latter case, given high--dimensional data $%
x_1,x_2,\ldots ,x_n$ in $\R^K$ for $K$ large compute a matrix of pairwise
distances, dist$(x_i,x_j)=D$. If classical Gower--Torgerson scaling is
applied to $D$, the result is essentially identical with principal
components analysis when used for dimension reduction.
We analyzed the oils and
fats data because this data set has been explained in the context of
Principal Components Analysis of interval data and therefore we can compare
our results with those obtained from Principal Components.

\begin{figure}[h]
\begin{center}
\includegraphics[height=8cm]{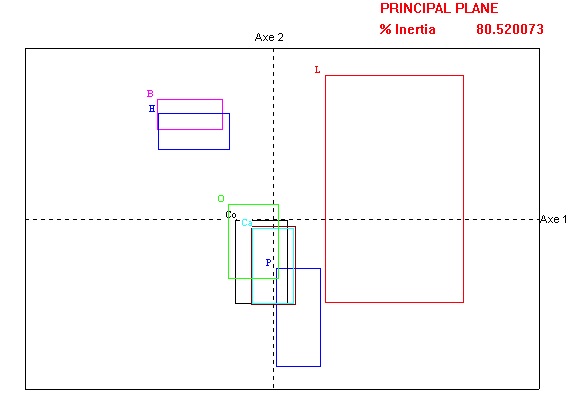}
\end{center}
\caption{Principal plane of oils and fats data with Multidimensional Scaling.}
\label{fig23}
\end{figure}

\begin{figure}[h]
\begin{center}
\includegraphics[height=8cm]{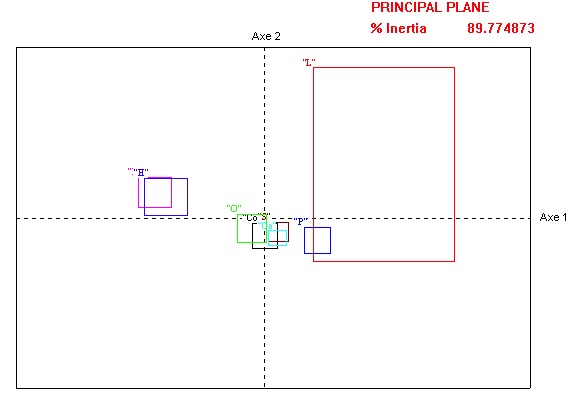}
\end{center}
\caption{Principal plane of oils and fats data using Principal Component
Analysis.}
\label{fig24}
\end{figure}

The oils and fats data set is shown in (Cazes, Chouakria, Diday and
Schektman (1997)). Each row of the data table refers to a class of oil
described by 4 quantitative interval type variables, ``Specific gravity'',
``Freezing point'', ``Iodine value'' and ``Saponification''. The matrix of
distances $\Delta $ that we used as an input to the INTERSCAL
multidimensional scaling was computed using a matrix $X$ that we got
standardizing the oils and fats matrix. To computed $\Delta $ we used the
equations (\ref{eq:eq211}) and (\ref{eq:eq212}). Using our INTERSCAL
algorithm we get the principal plane shown in Figure \ref{fig23}.
If we use Tops Principal Component Analysis with the oils and fats interval
data we get the results that are shown in Figure \ref{fig24}.

The clustering structure get in Figure \ref{fig23} and in Figure \ref{fig24}
are similar because the groups are the same and the size of the rectangles
are proportional. So the interpretation of both graphs will be just about
the same.

The second data set we considered consists of dissimilarity judgments of
rectangles of different area and height--to--width ratio, judged by 16
subjects. These data were presented in a paper on constrained
multidimensional scaling, (Winsberg and De Soete (1997)). Other researchers
have looked at rectangles. However, in general, they restricted their
attention to rectangles where the height is greater than the width or vice
versa. This data set includes both rectangles in which the height is greater
than the width and vice versa. In a study of rectangle dominance data
discussed by (Carroll (1972)) the consensus dimension corresponded fairly
well with size; but it was also clear in that case that subjects vary
greatly as to what they mean by size. Some subjects equated size to height,
some to area, some to width, and some to height--to--width ratio. When
Winsberg and De Soete (1997) analyzed their data for the 16 subjects, taken
together, three dimensions were recovered: the first was area, which relates
to size; the second dimension was height--to--width ratio, with recovered
values falling into essentially three categories, depending on whether the
height--to--width ratio was greater than, equal to, or less than one, which
relates to the position of the rectangle, (up--down); the third was
height--to--width ratio, or alternatively width--to--height ratio, such that the
value was less than or equal to one, i.e. squareness. So, the first
dimension relates to size, and the other two dimensions relate to shape.
Three latent classes were found in the CLASCAL analysis. The difference
among the classes was primarily due to how strongly they weighted dimension
two.

\begin{figure}[h]
\begin{center}
\includegraphics[height=8cm]{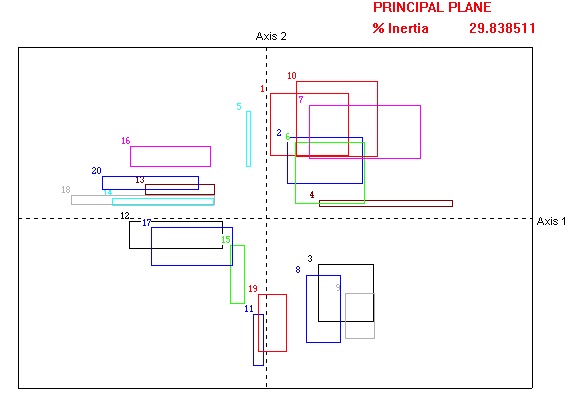}
\end{center}
\caption{Rectangles whose height is less than their width on the right.}
\label{fig241}
\end{figure}

\begin{figure}[h]
\begin{center}
\includegraphics[height=8cm]{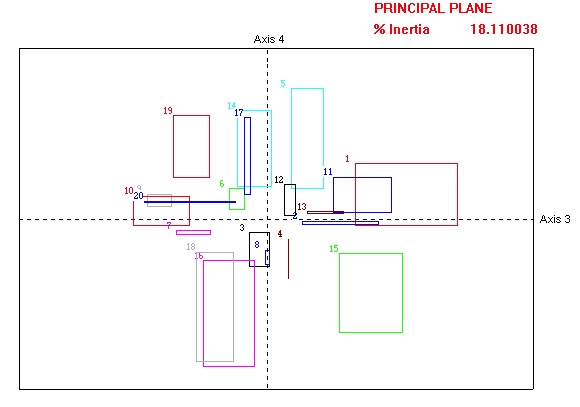}
\end{center}
\caption{Third dimension related to size or area with the smaller rectangles
appearing on the top.}
\label{fig242}
\end{figure}

Our INTERSCAL solution for the data recovers the same three dimensions.
Figures \ref{fig241} and \ref{fig242} display the results. The second dimension
separates the rectangles whose height is less than their width at the up
of Figure \ref{fig241} form those whose height is greater than their
width at the bottom of Figure \ref{fig241}. Dimension one is
related to squareness, that is width--to--height ratio or height--to--width ratio
whichever is less
than one. The rectangles which are nearly square are on the right side of Figure
\ref{fig241}. The third dimension is related to size or area with the
smaller rectangles appearing on the top of Figure \ref{fig242}.

Note that each stimulis/object is represented as a hypercube of
three dimensions. Thus for rectangle number eight we have $%
a(w)=[Y_1(w) \in [4.43,7.22]]\wedge [Y_2(w) \in [-35.12,-14.94]]\wedge
 [Y_3(w) \in [-0.05,0.49]]$. The
``psychological" rectangles occupy a hypercube so that for the physical
stimulis/object rectangle number eight, the model of the corresponding psychological object
is the symbolic object with a conjunction of three attributes, each
described by an interval, one interval for up--down $[4.43,7.22]$%
, one interval for squareness or shape $[-35.12,-14.94]$, (height--to--width
ratio or width--to--height ratio whichever is less than one), and one interval for area or
size $[-0.05,0.49]$. Note that up--down is not precisely localized. It is
represented by an interval for each symbolic object, even though the
``physical" rectangles fall into three categories on this variable that is,
up, (the height is greater than the width), down, (the width is greater than
the height), or neither, (the rectangle is a square). Up--down is not
precisely localized for each ``psychological" rectangle, because for some of
the judges, this dimension was more important than for others when making
the dissimilarity judgements, causing the distance between the up rectangles
and the down rectangles to be an interval. Note that the size of this
interval is smaller for those rectangles which are more nearly square, that
is those rectangles at the bottom of Figure \ref{fig241}.

These results are consistent with the results of the analyses presented in
Winsberg and De Soete (1997). In addition, this new technique indicates how
precisely the rectangles are located in the space. We have obtained the
interesting result that the size of the hypercube occupied by a
rectangle is inversely related to its area ($r=-0.72$). This finding
indicates that it is easier for subjects to distinguish larger rectangles
from one another than it is to do so for smaller rectangles.

\section{Conclusion}

We have presented a Multidimensional Scaling technique which enables the
representation of objects as hypercubes in a $n$ dimensional space,
reflecting a range of dissimilarities observed for each pair of objects. By
representing these objects as hypercubes we can display information relating
to how well the objects are localized. Moreover, as we demonstrated in the
example dealing with rectangles the precision with which an object is
localized may be related to one of its attributes.

This technique can be extended to include the case where in addition to the
common dimensions shared by all the stimuli, some stimuli may possess
specific attributes.

\begin{enumerate}
\item Bock, H-H and Diday, E. (eds.) (2000). {\em Analysis of Symbolic Data. Exploratory methods
for extracting statistical information from complex data.} Springer Verlag,
Heidelberg, 425 pages, ISBN 3-540-66619-2, 2000.

\item Borg I. and Groenen P. (1997). {\em Modern Multidimensional Scaling -- Theory and Applications},
Springer--Verlag, New York.

\item Brito P. (1991). {\em Analyse de donnees symboliques: Pyramides d'heritage},
Th\`ese de doctorat, Universit\'e Paris 9 Dauphine.

\item Carroll, J.D. (1972). {\em Individual Differences and Multidimensional Scaling}.
in  Multidimensional Scaling Theory and Applications in the
Behavioral Sciences, vol I, Theory, New York: Seminar Press.

\item Cazes P., Chouakria A., Diday E. et Schektman Y. (1997). Extension de l'analyse en composantes
principales \`a des donn\'ees de type intervalle, {\em Rev. Statistique Appliqu\'ee}, Vol. XLV Num. 3 pag. 5-24, France.

\item Cox T. and Cox M. (1994). {\em Multidimensional Scaling}, Chapman and Hall,New York.

\item Denoeux T. and Masson M. (1999). {\em Multidimensional Scaling of interval--valued dissimilarity data}.
Universit\'e de Technologie de Compi\'egne, France.

\item Diday E., Lemaire J., Pouget J., Testu F. (1984). {\em El\'ements d'Analyse des donn\'ees}. Dunod, Paris.

\item Diday E. (1987). {\em Introduction l'approche symbolique en Analyse des Donn\'es}. Premi\`ere Journ\'ees
Symbolique-Num\'erique. Universit\'e Paris IX Dauphine.

\item Gower, J. C. (1966) Some distances properties of latent root and vector methods
using multivariate analysis. {\em Biometrika}, 53, 325--338.

\item Torgenson, W. S. (1952) Multidimensional scaling: 1 Theory and method,
{\em Psychometrika}, 17, 401--419.

\item Torgenson, W. S. (1958) {\em Theory and methods of scaling}.
New York: Wiley.

\item Winsberg, S. and Desoete, G. (1997) Multidimensional scaling with constrained
dimensions: CONSCAL, {\em British Journal of Mathematical and Statistical Psychology}, 50, 55-72.
\end{enumerate}

\end{document}